\documentclass[aps,prd,reprint,superscriptaddress,longbibliography,showpacs,showkeys,nofootinbib,floatfix]{revtex4-2}

\usepackage[T1]{fontenc}
\usepackage[utf8]{inputenc}
\usepackage{amsmath,amssymb,mathtools,bm}
\usepackage{graphicx}
\usepackage{xcolor}
\usepackage{microtype}
\usepackage[colorlinks=true,linkcolor=blue!50!black,citecolor=blue!50!black,urlcolor=blue!50!black]{hyperref}

\newcommand{\dd}{\mathrm{d}}
\newcommand{\e}{\mathrm{e}}
\newcommand{\Ric}{\mathcal{R}}
\newcommand{\arsinh}{\operatorname{arsinh}}

\begin{document}

\title{A Survey through Conformal Time}

\author{T.~Aeenehvand}
\email{Taherehaeenehvand@gmail.com}
\affiliation{Theoretical Physics Group, Faculty of Physics, Alzahra University, Tehran, Iran}

\author{A.~Shariati}
\email{a.shariati@alzahra.ac.ir}
\affiliation{Theoretical Physics Group, Faculty of Physics, Alzahra University, Tehran, Iran}

\date{\today}

\begin{abstract}
We revisit conformal time $\eta$ in a spatially flat Friedmann--Robertson--Walker universe and use a $1+1$-dimensional setting as a technically transparent pedagogical arena. Our purpose is to clarify the relation among cosmic time $t$, conformal time $\eta$, and the scale factor $a(t)$, and then to follow how this relation governs the geodesics of freely moving particles and the curvature of the corresponding manifold. The radiation-dominated, matter-dominated, and exact vacuum-only de Sitter cases are treated separately, because each of them produces a distinct conformal-time dependence and therefore a distinct geodesic structure. We then write the affine-parameter formalism in a form that is genuinely general for any spatially flat conformal metric, and we record the straightforward extension to the spatially flat $3+1$ case. The presentation remains elementary in spirit, but the notation, the curvature formulas, and the de Sitter interpretation are kept explicit.
\end{abstract}


\maketitle

\section{Introduction}
Conformal time is often introduced in cosmology as a convenient reparametrization of the Friedmann--Robertson--Walker (FRW) metric, but its role is more than technical. In conformal form, the causal structure becomes especially transparent, which is why conformal coordinates recur in discussions of horizons, perturbations, quantum fields in curved spacetime, and geodesic structure~\cite{weinberg,hobson,mukhanov,dodelson,liddle,carroll,wald,birrelldavies,cunningham,bikwa,nemoul}. In that sense, conformal time is not merely an alternate clock; it is a geometric reorganization of the cosmological problem.

This geometric viewpoint also appears in recent discussions of time-reversal-symmetric and $CPT$-symmetric cosmologies, where the behavior of the scale factor near the Big Bang and across conformal descriptions of the background is conceptually important~\cite{robles,boyle}. At the same time, the term \emph{vacuum} must be used carefully. Here it refers only to a vacuum-energy source in classical Einstein--FRW dynamics, not to the broader quantum-gravitational notion of gravitational vacuum discussed by Mathur~\cite{mathur}. We also distinguish an \emph{exact} de Sitter spacetime---the vacuum solution with positive cosmological constant and no additional matter source, with $a(t)\propto \e^{Ht}$ in the spatially flat slicing---from a more general \emph{vacuum-dominated} era, which is only approximately or asymptotically de Sitter~\cite{spradlin,carrollcc,padmanabhancc}.

Our aim is geometric rather than encyclopedic. We track how conformal time reshapes the metric, the Christoffel symbols, the curvature, and the trajectories of freely moving particles. The analysis is carried out first in $1+1$ dimensions, where conformal flatness is especially transparent, and then briefly extended to the spatially flat $3+1$ case. The paper proceeds in three stages: first, the relation among $t$, $\eta$, and $a(\eta)$ is derived for radiation domination, matter domination, and the exact de Sitter limit; next, the geodesic equations, their first integrals, and the associated curvature are computed; finally, the affine-parameter formalism is written in a genuinely general way and extended to the spatially flat $3+1$ geometry. Throughout, we use the standard notation $a(t)$ for the scale factor to remain aligned with common cosmological usage.

\section{Conformal time in three benchmark cosmologies}
In $1+1$ dimensions, the spatially flat FRW metric may be written as
\begin{equation}
\dd s^2 = c^2\dd t^2-a^2(t)\dd x^2 = a^2(\eta)\left(\dd\eta^2-\dd x^2\right),
\label{eq:frwmetric}
\end{equation}
where conformal time is defined by
\begin{equation}
\dd\eta = \frac{c\,\dd t}{a(t)}.
\label{eq:defeta}
\end{equation}
We consider a power-law scale factor
\begin{equation}
a(t)=a_0\left(\frac{t}{t_0}\right)^{\alpha},\qquad \alpha\neq 1.
\label{eq:powerlaw}
\end{equation}
Integrating Eq.~\eqref{eq:defeta} gives
\begin{equation}
\eta(t)=\frac{c t_0}{a_0(1-\alpha)}\left(\frac{t}{t_0}\right)^{1-\alpha}+\eta_*,
\label{eq:etaoft}
\end{equation}
where $\eta_*$ is an integration constant. For $0<\alpha<1$ it is natural to choose $\eta_*=0$ so that $\eta\to 0$ as $t\to 0^+$. Inverting Eq.~\eqref{eq:etaoft} yields
\begin{align}
\frac{t}{t_0}&=\left[\frac{a_0(1-\alpha)}{c t_0}\,\eta\right]^{\!\frac{1}{1-\alpha}},\nonumber\\
a(\eta)&=a_0\left[\frac{a_0(1-\alpha)}{c t_0}\,\eta\right]^{\!\frac{\alpha}{1-\alpha}}.
\label{eq:aeta-general}
\end{align}
The case $\alpha=1$ is special and gives $\eta \propto \ln t $. 
\newpage
Three standard examples are:
\subsection{Radiation domination}
For a radiation-dominated universe, $a(t)\propto t^{1/2}$ so Eq.~\eqref{eq:aeta-general} becomes
\begin{equation}
a(\eta)=A_{\rm r}\,\eta,
\qquad
A_{\rm r}=\frac{a_0^2}{2ct_0}.
\label{eq:arad}
\end{equation}

\subsection{Matter domination}
For a matter-dominated universe, with $a(t)\propto t^{2/3}$, one obtains
\begin{equation}
a(\eta)=A_{\rm m}\,\eta^2,
\qquad
A_{\rm m}=\frac{a_0^3}{9c^2t_0^2},
\label{eq:amatter}
\end{equation}

\subsection{Vacuum domination}
For the exact de Sitter flat patch, corresponding to a positive cosmological constant with no additional matter component, the Hubble parameter is constant, $a(t)=\e^{Ht}.$ Integrating Eq.~\eqref{eq:defeta} and choosing the constant so that $\eta\to0^-$ as $t\to+\infty$, we obtain
\begin{equation}
a(\eta)=-\frac{\ell}{\eta},
\qquad
\ell\equiv\frac{c}{H},
\qquad \eta<0.
\label{eq:desitter-aeta}
\end{equation}

\begin{figure}[t]
    \centering
    \includegraphics[width=\columnwidth]{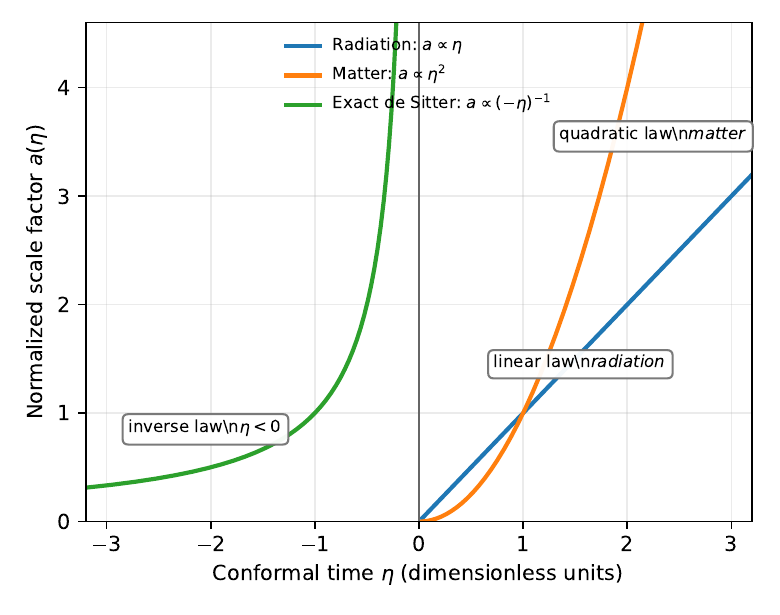}
    \caption{Scale factor as a function of conformal time for the three benchmark cosmologies studied in the text. The inset annotations are meant to keep the qualitative hierarchy visible at a glance: linear growth in the radiation case, quadratic growth in the matter case, and inverse behavior in the exact de Sitter limit.}
    \label{fig:scale-factor}
\end{figure}

Figure~\ref{fig:scale-factor} collects the three conformal-time scalings used throughout the rest of the paper. The annotation boxes in the figure summarize the only qualitative facts we need later: radiation gives a linear law, matter gives a quadratic law, and the exact de Sitter limit gives the inverse law $a\propto(-\eta)^{-1}$ on the negative-$\eta$ branch.

\section{Geodesics in $1+1$ dimensions}
For the metric \eqref{eq:frwmetric}, the only nonvanishing Christoffel symbols are
\begin{equation}
\Gamma^{\eta}_{\eta\eta}=\Gamma^{\eta}_{xx}=\Gamma^{x}_{\eta x}=\Gamma^{x}_{x\eta}=\frac{a'(\eta)}{a(\eta)},
\label{eq:christoffels}
\end{equation}
where the prime denotes differentiation with respect to $\eta$. The geodesic equations for an affine parameter, $\lambda$, therefore read
\begin{align}
\eta''+\frac{a'}{a}\left(\eta'^2+x'^2\right)&=0,
\label{eq:geo-eta}\\
 x''+2\frac{a'}{a}\eta' x'&=0.
\label{eq:geo-x}
\end{align}
Because $x$ is a cyclic coordinate, there is an immediate first integral,
\begin{equation}
p\equiv a^2(\eta)x'=\text{constant}.
\label{eq:pconst}
\end{equation}
A second first integral comes from the norm of the tangent vector,
\begin{align}
a^2(\eta)\left(\eta'^2-x'^2\right)&=\kappa,\nonumber\\
\kappa&=\begin{cases}
+1,& \text{timelike geodesic},\\
0,& \text{null geodesic},\\
-1,& \text{spacelike geodesic}.
\end{cases}
\label{eq:normconst}
\end{align}
Combining Eqs.~\eqref{eq:pconst} and \eqref{eq:normconst} gives
\begin{equation}
\eta'^2=\frac{p^2+\kappa a^2(\eta)}{a^4(\eta)},
\qquad
\frac{\dd x}{\dd\eta}=\frac{p}{\sqrt{p^2+\kappa a^2(\eta)}}.
\label{eq:first-integrals}
\end{equation}

\subsection{Null geodesics}
For $\kappa=0$, \eqref{eq:first-integrals} immediately gives
\begin{equation}
\frac{\dd x}{\dd\eta}=\pm 1,
\qquad x(\eta)=\pm\eta+x_0,
\label{eq:null}
\end{equation}
which is the familiar statement that causal structure is Minkowskian in conformal coordinates.

\subsection{Timelike geodesics in the radiation era}
\begin{figure}[t]
    \centering
    \includegraphics[width=\columnwidth]{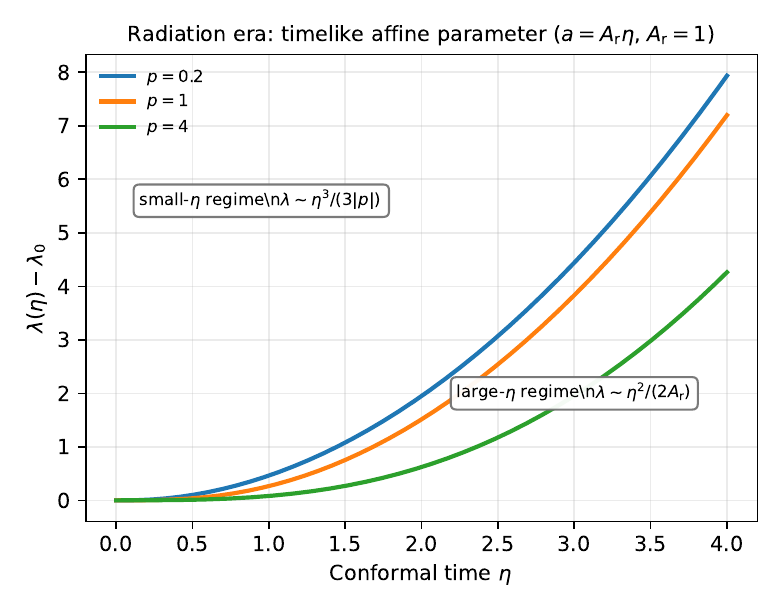}
    \caption{Exact affine parameter $\lambda(\eta)$ for timelike geodesics in the radiation era, generated from Eq.~\eqref{eq:lambda-rad} in dimensionless units with $A_{\rm r}=1$. The boxed notes reproduce the two asymptotic limits used in the text. Larger conserved momentum delays the transition between them.}
    \label{fig:affine-radiation}
\end{figure}

For radiation domination, $a(\eta)=A_{\rm r}\eta$, and Eq.~\eqref{eq:first-integrals} becomes
\begin{equation}
\frac{\dd\lambda}{\dd\eta}=\frac{A_{\rm r}^2\eta^2}{\sqrt{p^2+A_{\rm r}^2\eta^2}},
\qquad
\frac{\dd x}{\dd\eta}=\frac{p}{\sqrt{p^2+A_{\rm r}^2\eta^2}}.
\label{eq:rad-firstintegrals}
\end{equation}
Integrating gives
\begin{align}
\lambda-\lambda_0 &= \frac{\eta\sqrt{p^2+A_{\rm r}^2\eta^2}}{2}
\,\frac{1}{A_{\rm r}^2}
-\frac{p^2}{2A_{\rm r}^3}
\arsinh\!\left(\frac{A_{\rm r}\eta}{|p|}\right),
\label{eq:lambda-rad}\\
x-x_0 &= \frac{1}{A_{\rm r}}
\arsinh\!\left(\frac{A_{\rm r}\eta}{p}\right)
\quad (p\neq 0).
\label{eq:x-rad}
\end{align}
The small-$\eta$ behaviour is cubic,
\begin{equation}
\lambda-\lambda_0 \sim \frac{\eta^3}{3|p|}
\qquad (A_{\rm r}\eta\ll |p|),
\label{eq:small-eta-rad}
\end{equation}
while the large-$\eta$ behaviour becomes quadratic,
\begin{equation}
\lambda-\lambda_0 \sim \frac{\eta^2}{2A_{\rm r}}
\qquad (A_{\rm r}\eta\gg |p|).
\label{eq:large-eta-rad}
\end{equation}
Thus the affine parametrization interpolates smoothly between a momentum-dominated regime and a scale-factor-dominated regime. This is exactly the crossover emphasized in the annotation boxes of Fig.~\ref{fig:affine-radiation}: at early conformal times the conserved comoving momentum controls the affine evolution, whereas at larger $\eta$ the scale factor itself takes over.

\subsection{Timelike geodesics in the matter era}
\begin{figure}[t]
    \centering
    \includegraphics[width=\columnwidth]{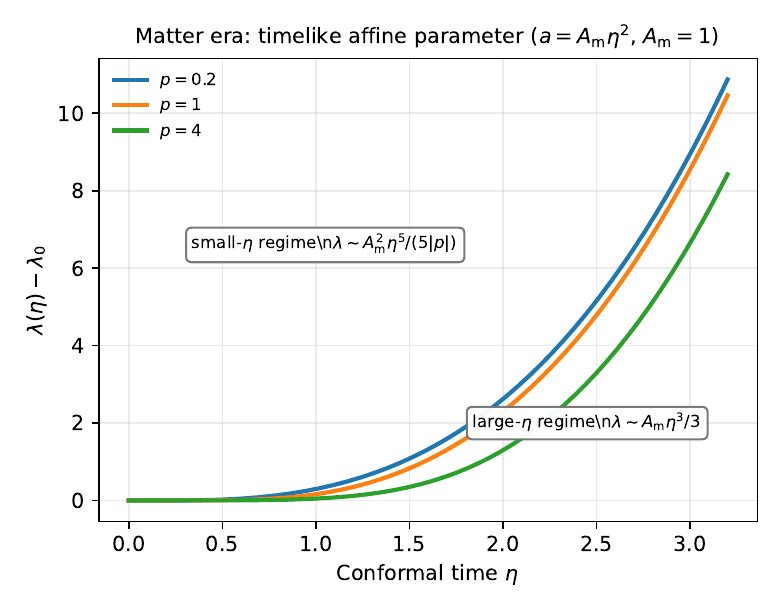}
    \caption{Numerical affine parameter $\lambda(\eta)$ for timelike geodesics in the matter era, obtained from Eq.~\eqref{eq:matter-firstintegrals} in dimensionless units with $A_{\rm m}=1$. The two boxed asymptotic laws emphasize the same momentum-to-expansion crossover as in the radiation case, now with quintic early-time and cubic late-time growth.}
    \label{fig:affine-matter}
\end{figure}

For matter domination, $a(\eta)=A_{\rm m}\eta^2$, and Eq.~\eqref{eq:first-integrals} gives
\begin{equation}
\frac{\dd\lambda}{\dd\eta}=\frac{A_{\rm m}^2\eta^4}{\sqrt{p^2+A_{\rm m}^2\eta^4}},
\qquad
\frac{\dd x}{\dd\eta}=\frac{p}{\sqrt{p^2+A_{\rm m}^2\eta^4}}.
\label{eq:matter-firstintegrals}
\end{equation}
In this case the exact antiderivatives are no longer elementary, but the asymptotic structure is still transparent and already captures the physical crossover. For $A_{\rm m}\eta^2\ll |p|$ one finds
\begin{equation}
\lambda-\lambda_0 \sim \frac{A_{\rm m}^2}{5|p|}\,\eta^5,
\label{eq:small-eta-mat}
\end{equation}
whereas for $A_{\rm m}\eta^2\gg |p|$ one obtains
\begin{equation}
\lambda-\lambda_0 \sim \frac{A_{\rm m}}{3}\,\eta^3.
\label{eq:large-eta-mat}
\end{equation}
The corresponding conformal-time scale is set by
\begin{equation}
A_{\rm m}\eta_c^2\sim |p|,
\qquad
\eta_c\sim \sqrt{\frac{|p|}{A_{\rm m}}}.
\label{eq:etac-matter}
\end{equation}
Thus the matter era exhibits the same momentum-to-expansion transition as the radiation era, but with different powers because the quadratic conformal growth of the scale factor stretches the affine parameter more strongly at late times. The spatial motion reflects the same pattern: for small $\eta$ the trajectory is nearly null, $x-x_0\sim \operatorname{sgn}(p)\eta$, whereas for large $\eta$ one has $\dd x/\dd\eta\sim p/(A_{\rm m}\eta^2)$, so the comoving position approaches a finite asymptote. A numerical illustration of the corresponding affine histories is given later in Fig.~\ref{fig:affine-matter}.

\subsection{Timelike geodesics in de Sitter spacetime}
\begin{figure}[t]
    \centering
    \includegraphics[width=0.98\columnwidth]{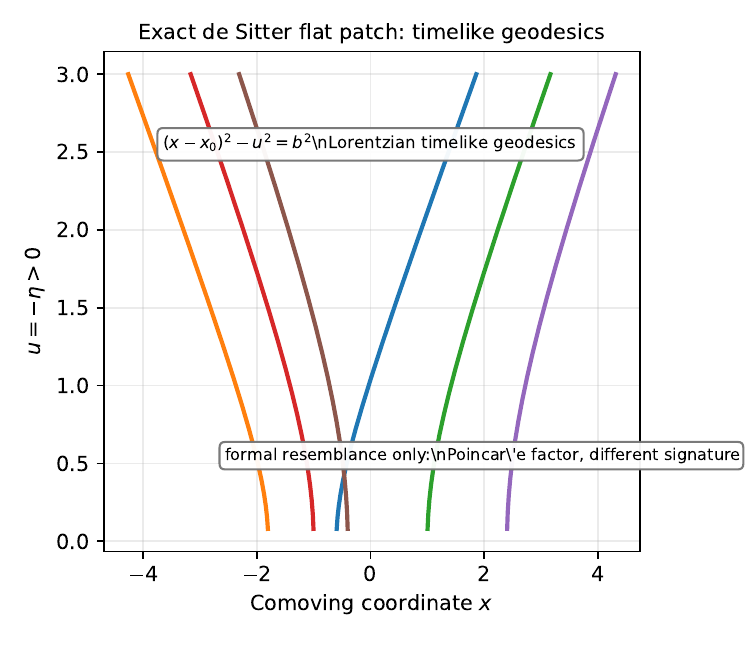}
    \caption{Families of timelike geodesics in the $1+1$ exact de Sitter flat patch, written in terms of $u=-\eta>0$. Each branch satisfies $(x-x_0)^2-u^2=b^2$ with $b=\ell/|p|$. The annotation box emphasizes that these hyperbolae follow directly from the Lorentzian geodesic equations.}
    \label{fig:desitter-geodesics}
\end{figure}

For de Sitter expansion, it is convenient to define $u\equiv -\eta>0$, so that
\begin{equation}
\dd s^2 = \frac{\ell^2}{u^2}\left(\dd u^2-\dd x^2\right).
\label{eq:desitter-u}
\end{equation}
For timelike geodesics ($\kappa=1$), Eq.~\eqref{eq:first-integrals} yields
\begin{equation}
\frac{\dd x}{\dd u}=\pm\frac{p u}{\sqrt{\ell^2+p^2u^2}}.
\label{eq:dxdu-ds}
\end{equation}
The integral is elementary:
\begin{equation}
x-x_0 = \pm\frac{1}{p}\sqrt{\ell^2+p^2u^2},
\label{eq:dS-trajectory}
\end{equation}
which can be rewritten as
\begin{equation}
(x-x_0)^2-u^2=b^2,
\qquad
b\equiv\frac{\ell}{|p|}.
\label{eq:dS-hyperbola}
\end{equation}
Therefore timelike geodesics in the $(x,u)$ plane are hyperbolae. This result supplies a direct Lorentzian derivation of the hyperbolic trajectories which is precisely underlined in the boxed statement in Fig.~\ref{fig:desitter-geodesics}.

\section{Curvature and the conformal-factor analogy}
With the sign convention
\begin{equation}
R^{\rho}{}_{\sigma\mu\nu} = \partial_{\mu}\Gamma^{\rho}_{\nu\sigma}-\partial_{\nu}\Gamma^{\rho}_{\mu\sigma}
+\Gamma^{\rho}_{\mu\lambda}\Gamma^{\lambda}_{\nu\sigma}-\Gamma^{\rho}_{\nu\lambda}\Gamma^{\lambda}_{\mu\sigma},
\label{eq:riemann-sign}
\end{equation}
the independent Riemann component and the Ricci scalar are
\begin{equation}
R_{\eta x\eta x}=a a''-a'^2,
\qquad
\Ric = \frac{2\left(a'^2-a a''\right)}{a^4} = -\frac{2}{c^2}\frac{\ddot a}{a}.
\label{eq:ricci-general}
\end{equation}
The final equality follows after transforming back to cosmic time.

For the three benchmark cosmologies one obtains
\begin{align}
\text{radiation:}\qquad & \Ric = \frac{2}{A_{\rm r}^2\eta^4},
\label{eq:Ric-rad}\\
\text{matter:}\qquad & \Ric = \frac{4}{A_{\rm m}^2\eta^6},
\label{eq:Ric-mat}\\
\text{de Sitter:}\qquad & \Ric = -\frac{2}{\ell^2}=-\frac{2H^2}{c^2}.
\label{eq:Ric-dS}
\end{align}
Thus the radiation- and matter-dominated $1+1$ geometries are not Ricci-flat, and the de Sitter scalar curvature is constant and maximally symmetric as expected. With the opposite Riemann-tensor sign convention, the constant de Sitter value in Eq.~\eqref{eq:Ric-dS} would appear with the opposite sign.

It is tempting to compare Eq.~\eqref{eq:desitter-u} with the Poincar\'e half-plane metric,
\begin{equation}
\dd s^2_{\rm H} = \frac{\ell^2}{u^2}\left(\dd u^2+\dd x^2\right).
\label{eq:half-plane}
\end{equation}
The conformal factor is indeed the same, but the signature is not. Therefore the relation is only formal: after a Wick rotation or a sign flip one recovers the Euclidean half-plane, but the Lorentzian de Sitter flat patch is not isometric to it, and the geodesic families should not be identified blindly~\cite{anderson,stahl}.

\section{Extension to spatially flat $3+1$ FRW spacetime}
For the spatially flat $3+1$ metric in Cartesian comoving coordinates,
\begin{equation}
\dd s^2 = a^2(\eta)\left(\dd\eta^2-\dd\bm{x}^{2}\right),
\qquad \bm{x}=(x^1,x^2,x^3),
\label{eq:4dmetric}
\end{equation}
spatial translation invariance gives three conserved comoving momenta,
\begin{equation}
P_i = a^2(\eta)\,\frac{\dd x^i}{\dd\lambda},
\qquad i=1,2,3.
\label{eq:Pi}
\end{equation}
With $P^2\equiv\delta^{ij}P_iP_j$, the normalization condition becomes
\begin{equation}
a^2(\eta)\left(\eta'^2-\delta_{ij}x^{i\prime}x^{j\prime}\right)=\kappa,
\label{eq:4dnorm}
\end{equation}
so that
\begin{equation}
\eta'^2 = \frac{P^2+\kappa a^2(\eta)}{a^4(\eta)},
\qquad
\frac{\dd x^i}{\dd\eta}=\frac{P_i}{\sqrt{P^2+\kappa a^2(\eta)}}.
\label{eq:4dfirstintegrals}
\end{equation}
The comoving spatial trajectory is therefore a straight line,
\begin{equation}
\bm{x}(\eta)=\bm{x}_0+\hat{\bm{P}}\int_{\eta_0}^{\eta}\frac{P\,\dd\tilde\eta}{\sqrt{P^2+\kappa a^2(\tilde\eta)}},
\qquad
\hat{\bm{P}}\equiv \frac{\bm{P}}{P},
\label{eq:4dstraightline}
\end{equation}
with nontrivial parametrization encoded entirely in the scale factor.

If one instead works in spherical comoving coordinates and restricts the motion to the equatorial plane, then the angular momentum
\begin{equation}
J = a^2(\eta)r^2\phi'
\label{eq:angularmomentum}
\end{equation}
is conserved, and the radial equation becomes
\begin{equation}
r'^2 = \eta'^2-\frac{\kappa}{a^2(\eta)}-\frac{J^2}{a^4(\eta)r^2}.
\label{eq:radialequation}
\end{equation}
Equation \eqref{eq:radialequation} is the natural $3+1$ generalization of the $1+1$ analysis when angular motion is present.

\section{Affine parameter and physical interpretation}
The general affine-parameter statement can now be written without over-specializing to any one era. For a spatially flat conformal metric of the form \eqref{eq:frwmetric},
Eqs.~\eqref{eq:pconst} and \eqref{eq:normconst} imply
\begin{equation}
\left(\frac{\dd\eta}{\dd\lambda}\right)^2=\frac{p^2+\kappa a^2(\eta)}{a^4(\eta)}.
\label{eq:etaaffine}
\end{equation}
Therefore,
\begin{equation}
\lambda-\lambda_0=\pm\int \frac{a^2(\eta)\,\dd\eta}{\sqrt{p^2+\kappa a^2(\eta)}}.
\label{eq:general-affine}
\end{equation}
Equation~\eqref{eq:general-affine} is the proper general formalism. Whether it is elementary or not depends on the explicit form of $a(\eta)$.

For timelike geodesics one may choose the affine parameter to be the arc length, $\lambda=s=c\tau$, where $\tau$ is the particle proper time. In that normalization the relation between $\lambda$ and conformal time says directly how the particle's own clock is related to $\eta$. In particular, for a comoving observer with $p=0$ one has $\dd\lambda=a(\eta)\,\dd\eta=c\,\dd t$, so the affine parameter reduces to cosmic proper time up to the factor $c$. This makes the distinction between $t$ and $\eta$ especially explicit: conformal time is not itself the physical time measured by a freely falling comoving clock, but becomes so only after multiplication by the scale factor.

A second interpretation is obtained by using the orthonormal frame carried by comoving observers,
\begin{equation}
e_{\hat 0}=a^{-1}\partial_{\eta},
\qquad
 e_{\hat x}=a^{-1}\partial_x.
\label{eq:orthonormal-frame}
\end{equation}
For timelike geodesics with $\lambda=s$, the frame components of the four-velocity are
\begin{equation}
u_{\hat 0}=a\frac{\dd\eta}{\dd s}=\frac{\sqrt{p^2+a^2(\eta)}}{a(\eta)},
\qquad
u_{\hat x}=a\frac{\dd x}{\dd s}=\frac{p}{a(\eta)}.
\label{eq:frame-velocity}
\end{equation}
Hence the peculiar velocity measured by comoving observers is
\begin{equation}
v_{\rm pec}=\frac{u_{\hat x}}{u_{\hat 0}}=\frac{p}{\sqrt{p^2+a^2(\eta)}},
\label{eq:peculiar-velocity}
\end{equation}
in units of $c$. The conserved quantity $p$ can therefore be read as the comoving momentum per unit rest mass, while the corresponding physical peculiar momentum redshifts as $a^{-1}$. When $a(\eta)\ll |p|$, the motion is momentum-dominated and the particle is strongly non-comoving. When $a(\eta)\gg |p|$, one has $v_{\rm pec}\approx p/a(\eta)\to 0$, so the particle is gradually carried into the Hubble flow.

The null case is conceptually different. A photon has no proper time, but it still admits an affine parameter. Setting $\kappa=0$ in Eq.~\eqref{eq:general-affine} gives
\begin{equation}
\dd\lambda=\pm\frac{a^2(\eta)}{|p|}\,\dd\eta.
\label{eq:null-affine}
\end{equation}
Thus null trajectories remain straight lines in the $(\eta,x)$ plane, but equal increments of conformal time do not correspond to equal affine intervals. In this sense the conformal diagram hides part of the physical evolution: the path looks Minkowskian, while the affine stretching still remembers the expansion. The same mechanism underlies the familiar cosmological redshift law for photon energies measured by comoving observers.

In the radiation case, where $a(\eta)=A_{\rm r}\eta$, Eq.~\eqref{eq:general-affine} reduces to Eq.~\eqref{eq:lambda-rad}. The transition between the cubic and quadratic regimes occurs when the two terms in the square root are comparable, namely when
\begin{equation}
A_{\rm r}^2\eta_c^2\sim p^2,
\qquad \text{or equivalently} \qquad
\eta_c\sim \frac{|p|}{A_{\rm r}}.
\label{eq:etac}
\end{equation}
This critical conformal-time scale marks the point at which the conserved comoving momentum ceases to dominate the affine evolution.

For the matter law $a(\eta)=A_{\rm m}\eta^2$, the same momentum-to-expansion crossover persists, but the asymptotic powers change to
\begin{align}
\lambda-\lambda_0 &\sim \frac{A_{\rm m}^2}{5|p|}\,\eta^5,
&& A_{\rm m}\eta^2\ll |p|,
\label{eq:matter-asymp-1}\\
\lambda-\lambda_0 &\sim \frac{A_{\rm m}}{3}\,\eta^3,
&& A_{\rm m}\eta^2\gg |p|.
\label{eq:matter-asymp-2}
\end{align}
The crossover scale $a(\eta_c)\sim |p|$ therefore has a direct physical meaning: it marks the epoch at which peculiar motion and cosmological expansion contribute comparably to the normalization condition. Before that epoch the affine evolution is controlled mainly by the conserved comoving momentum; after it the affine history follows the background expansion more directly.

From a geometric point of view, the content of Eq.~\eqref{eq:general-affine} is simple. The affine parameter is not determined by conformal time alone; it is determined by conformal time together with the scale factor and the conserved spatial momentum. The same conformal coordinate can therefore correspond to different affine histories in different cosmological backgrounds. That distinction is one of the reasons the conformal-time description, although very clean geometrically, still remembers the matter content of the universe in a nontrivial way.

\section{Conclusion}
Conformal time simplifies the FRW metric, but it does not erase the physical imprint of the cosmic medium. Radiation domination, matter domination, and vacuum domination lead to different scale factors in conformal coordinates, and these differences propagate directly into the geodesic equations, the affine parametrization of particle motion, and the curvature of the underlying manifold. Within the $1+1$ model, null geodesics remain straight in conformal coordinates, whereas timelike geodesics retain a memory of the background through the conformal factor. In the radiation era this yields an exact affine-parameter expression with a clean crossover between momentum-dominated and scale-factor-dominated regimes; in the matter era the same crossover persists with distinct asymptotic powers and a stronger late-time affine stretching; and in the exact de Sitter patch the scalar curvature is constant and the timelike trajectories in the $(x,-\eta)$ plane are hyperbolae. Read in that way, the examples form a compact pedagogical comparison: the conformal diagram makes the causal structure simple, while the affine parameter and curvature keep track of the physical background.

A second conclusion concerns interpretation. Treating the exact de Sitter solution and a generic vacuum-dominated era as interchangeable is physically imprecise. The former is an exact vacuum solution with $a(t)=\e^{Ht}$, whereas the latter is usually an asymptotic or approximate late-time regime. It is equally important to remember that the $1+1$ treatment is a kinematical guide, not a full replacement for the Einstein-dynamical content of spatially flat $3+1$ cosmology. The brief extension to spatially flat $3+1$ spacetime shows why the simpler model is still useful: the conserved-quantity method survives almost unchanged, so the lower-dimensional discussion isolates the geometric mechanism without pretending to exhaust the realistic dynamics. A natural next step would be to admit mixtures of matter components, nonzero spatial curvature, or perturbative departures from exact flatness. Those generalizations would complicate the algebra, but they would not alter the central lesson of this note: conformal time is most informative when it is used not as a substitute for cosmic dynamics, but as a coordinate framework that makes those dynamics easier to read.
\begin{acknowledgments}
This research was carried out within the Theoretical Physics Group, Faculty of Physics, Alzahra University, Tehran, Iran. We also wish to thank Professor M. Khorrami for generous guidance and encouragement during the maturation of the project.
\end{acknowledgments}

\appendix
\section{General $1+1$ conformal factor $f(\eta,x)$}
For the more general metric
\begin{equation}
\dd s^2 = f^2(\eta,x)\left(\dd\eta^2-\dd x^2\right),
\label{eq:general-fmetric}
\end{equation}
the nonzero Christoffel symbols are
\begin{align}
\Gamma^{\eta}_{\eta\eta}=\Gamma^{\eta}_{xx}=\Gamma^{x}_{\eta x}=\Gamma^{x}_{x\eta} &= \frac{\partial_\eta f}{f},
\label{eq:genGamma1}\\
\Gamma^{\eta}_{\eta x}=\Gamma^{\eta}_{x\eta}=\Gamma^{x}_{\eta\eta}=\Gamma^{x}_{xx} &= \frac{\partial_x f}{f}.
\label{eq:genGamma2}
\end{align}
The geodesic equations become
\begin{align}
\eta'' + \frac{1}{f}\Big[\partial_\eta f\left(\eta'^2+x'^2\right)+2\partial_x f\,\eta' x'\Big] &=0,
\label{eq:genGeo1}\\
x'' + \frac{1}{f}\Big[\partial_x f\left(\eta'^2+x'^2\right)+2\partial_\eta f\,\eta' x'\Big] &=0.
\label{eq:genGeo2}
\end{align}
The independent Riemann component and the Ricci scalar are
\begin{align}
R_{\eta x\eta x} &= f\left(\partial^2_{\eta}f-\partial^2_x f\right)-\left(\partial_\eta f\right)^2+\left(\partial_x f\right)^2,
\label{eq:general-riemann}\\
\Ric &= \frac{2}{f^4}\Big[\left(\partial_\eta f\right)^2-\left(\partial_x f\right)^2-f\left(\partial^2_{\eta}f-\partial^2_x f\right)\Big].
\label{eq:general-ricci}
\end{align}
Equations \eqref{eq:general-riemann} and \eqref{eq:general-ricci} reduce to Eq.~\eqref{eq:ricci-general} when $f(\eta,x)=a(\eta)$.
\bibliographystyle{apsrev4-1}
\bibliography{conformal_time_refs}

\end{document}